\documentstyle[aps,epsf,rotate]{revtex}
\font\mb=msbm10

\newcommand{\bml}[1]{\mbox{{\boldmath$#1$}}}
\newcommand{\tl}[1]{\tilde{#1}}
\newcommand{\bc}{\begin{center}}
\newcommand{\ec}{\end{center}}
\newcommand{\be}{\begin{equation}}
\newcommand{\ee}{\end{equation}}
\newcommand{\ba}{\begin{eqnarray*}}
\newcommand{\ea}{\end{eqnarray*}}
\newcommand{\bna}{\begin{eqnarray}}
\newcommand{\ena}{\end{eqnarray}}
\begin{document}

\draft
\title{Understanding deterministic diffusion by correlated random walks}
\author{R.Klages$^{1}$ and N.Korabel$^2$}
\address{Max Planck Institute for Physics of Complex Systems,
N\"othnitzer Str. 38,  D-01187 Dresden, Germany\\
$^1$E-mail: rklages@mpipks-dresden.mpg.de\\
$^2$ E-mail: korabel@mpipks-dresden.mpg.de}
\date{\today}
\maketitle
\begin{abstract}
Low-dimensional periodic arrays of scatterers with a moving point particle are
ideal models for studying deterministic diffusion. For such systems the
diffusion coefficient is typically an irregular function under variation of a
control parameter. Here we propose a systematic scheme of how to approximate
deterministic diffusion coefficients of this kind in terms of correlated
random walks. We apply this approach to two simple examples which are a
one-dimensional map on the line and the periodic Lorentz gas. Starting from
suitable Green-Kubo formulas we evaluate hierarchies of approximations for
their parameter-dependent diffusion coefficients. These approximations
converge exactly yielding a straightforward interpretation of the structure of
these irregular diffusion coeficients in terms of dynamical correlations.
\end{abstract}

\section{Introduction}
Deterministic diffusion is a prominent topic in the theory of chaotic
dynamical systems and in nonequilibrium statistical mechanics
\cite{Do99,Gasp}. To achieve a proper understanding of the mechanism of
deterministic diffusion with respect to microscopic chaos in the equations of
motion, a number of simple model systems was proposed and analyzed. An
interesting finding was that, for a simple one-dimensional chaotic map on the
line, the diffusion coefficient is a fractal function of a control parameter
\cite{RKD,RKdiss,KlDo99}. Diffusion coefficients exhibiting similar
irregularities are also known for more complicated sytems such as sawtooth maps
\cite{DMP89}, standard maps \cite{ReWi80}, and Harper maps \cite{Lebo98}, the
latter two models being closely related to physical systems like the kicked
rotor, or a particle moving in a periodic potential under the influence of
electric and magnetic fields. More recently, related irregularities in the
diffusion coefficient were discovered in Hamiltonian billiards such as the
periodic Lorentz gas \cite{KlDe00}, and for a particle moving on a corrugated
floor under the influence of an external field \cite{HaGa01}. That this
phenomenon is not specific to diffusion coefficients is suggested by results
on other transport coefficients such as chemical reaction rates in multibaker
maps \cite{GaKl}, electrical conductivities in the driven periodic Lorentz gas
\cite{MH87}, and magnetoresistances in antidot lattices
\cite{Weis91} which, again, exhibit strongly irregular behavior under
parameter variation.

In this paper we wish to contribute to the further analysis and understanding
of irregular transport coefficients in simple model systems. We use certain
forms of the Green-Kubo formula as starting points for expansions of the
diffusion coefficient in terms of correlated random walks. In Section 2 we
exemplify this approach for the piecewise linear map on the line studied in
Refs.\ \cite{RKD,RKdiss,KlDo99}. In Section 3 we apply a suitably adapted
version of it to diffusion in the periodic Lorentz gas. In both cases our
approximations converge quickly to the precise diffusion coefficient as
obtained from other numerical methods. We argue that for both models this
approximation scheme provides a physical explanation for the irregular
structure of these diffusion coefficients in terms of dynamical correlations,
or memory effects. In Section 4 we summarize our results and relate them to
previous work on irregular and fractal transport coefficients.

\section{Deterministic diffusion in one-dimensional maps on the line}
Probably the most simple models exhibiting deterministic diffusion are
one-dimensional maps defined by the equation of motion
\begin{equation}
x_{n+1}=M_a(x_n) \label{eq:eom}
\end{equation}
where $a\in \hbox{\mb R}$ is a control parameter and $x_n$ is the position of
a point particle at discrete time $n$. $M_a(x)$ is continued periodically
beyond the interval $[0,1)$ onto the real line by a lift of degree one,
$M_a(x+1)=M_a(x)+1$. We assume that $M_a(x)$ is anti-symmetric with respect to
$x=0$, $M_a(x)=-M_a(-x)$. Let
\be
m_a(x):=M_a(x) \quad \mbox{mod} \; 1 \label{eq:mod}
\ee
be the reduced map related to $M_a(x)$. This map governs the dynamics on the
unit interval according to $x_n=m_a^n(x)$. Let $\rho_n(x)$ be the probability
density on the unit interval of an ensemble of moving particles starting at
initial conditions $x\equiv x_0$. This density evolves according to the
Frobenius-Perron continuity equation
\be
\rho_{n+1}(x)=\int_0^1 dy \: \rho_n(y) \:\delta(x-m_a(y)) \quad . \label{eq:fp} 
\ee
Here we are interested in the deterministic diffusion coefficient defined by
the Einstein formula
\be
D(a)=\lim_{n\to\infty}<(x_n-x_0)^2>/(2n) \label{eq:dkdef1}
\ee
with $x_n$ governed by $M_a(x)$, where the brackets denote an average over the
initial values $x_0$ with respect to the invariant probability density on the
unit interval, $\left\langle \ldots \right\rangle
:=\int_0^1dx\;\rho_a^*(x)\ldots$. Eq.(\ref{eq:dkdef1}) can be transformed onto
the Green-Kubo formula for maps
\cite{Do99,RKdiss,GF2,PG1},
\be
D(a) = \frac {1}{2}\left\langle j^2(x_0)\right\rangle
+\sum_{k=1}^\infty\left\langle j(x_0)j(x_k)\right\rangle
\quad , \label{eq:gk1}
\ee
where the jump velocity 
\be
j(x_n):=[x_{n+1}]-[x_n] \quad , \label{eq:jv}
\ee
$[x]$ being the largest integer less than $x$, takes only integer values and
denotes how many unit intervals a particle has traversed after one iteration
starting at $x_n$. The map we study as an example is defined by
\be
M_a(x)  = \left\{
\begin{array}{r@{\quad,\quad}l}
a x & 0< x\le\frac{1}{2} \\
a x +1-a & \frac{1}{2} < x\le 1
\end{array}
\right\}  \label{eq:mapa} \quad ,
\ee
where the uniform slope $a>2$ serves as a control parameter. The Lyapunov
exponent of this map is given by $\lambda(a)=\ln a$ implying that the dynamics
is chaotic. A sketch of this map is shown in Fig.\ \ref{fig1} (a). Fig.\
\ref{fig1} furthermore depicts the parameter-dependent diffusion coefficient
of this map in the regime of $2\le a\le4$ as calculated in Refs.\
\cite{RKD,RKdiss,KlDo99} by means of a numerical implementation of analytical
methods. In these references numerical evidence was provided that the
irregularities on a fine scale are reminiscent of an underlying fractal
structure.

To analyze this fractal diffusion coefficient, Eq.(\ref{eq:gk1}) forms a
suitable starting point because it distinguishes between two crucial
contributions of the dynamics to the diffusion process, which are (1) the
motion of the particle on the unit interval $x_n \;\mbox{mod}\; 1$ generating
the invariant density $\rho_a^*(x)$, and (2) the integer jumps from one unit
interval to another one as related to $j(x_n)$. In fact, in Refs.\
\cite{RKdiss,GaKl} it was shown that both parts are independent sources of
fractality for the diffusion coefficient. However, in these references Eq.\
(\ref{eq:gk1}) was only discussed in the limit of infinite time. In this paper
we suggest to look at the contributions of the single terms in this expansion,
and to analyze how the Green-Kubo formula approaches the exact diffusion
coefficient step by step. We do this by systematically building up hierarchies
of approximate diffusion coefficients. These approximations should be defined
such that the different dynamical contributions to the diffusion process are
properly filtered out. Another issue is how to evaluate the single terms of
the Green-Kubo expansion on an analytical basis. In this section we show how
to do this for the one-dimensional model introduced above, in the next section
we apply the same idea to billiards such as the periodic Lorentz gas.

We start by looking at the first term in Eq.(\ref{eq:gk1}). For $a<4$ the
absolute value of the jump velocity $j(x_n)$ is either zero or one.  Assuming
that $\rho_a^*(x)\simeq1$ for $a\to2$ and cutting off all higher order-terms
in Eq.\ (\ref{eq:gk1}), the first term leads to the well-known {\em random
walk approximation} of the diffusion coefficient
\cite{RKdiss,GF2,GeNi82,SFK,dcrc}
\be
D_0(a)=\frac {1}{2}\int_0^1dx\;j^2_a(x) \label{eq:dkrw1}
\ee
which in case of the map Eq.\ (\ref{eq:mapa}) reads
\be
D_0(a)=(a-2)/(2a) \quad . \label{eq:dkrw2}
\ee
This solution is asymptotically correct in the limit of $a\to2$. More
generally speaking, the reduction of the Green-Kubo formula to the first term
only is an exact solution for arbitrary parameter values only if all
higher-order contributions from the velocity autocorrelation function
$C(n):=<j(x_0)j(x_n)>$ are strictly zero. This is only true for systems of
Bernoulli type \cite{GF2}. Conversely, the series expansion in form of Eq.\
(\ref{eq:gk1}) systematically gives access to higher-order corrections by
including higher-order correlations, or memory effects. This leads us to the
definition of two hierarchies of correlated random walk diffusion
coefficients:

(1) Again, we make the approximation that $\rho_a^*(x)\simeq1$. We
then define
\be
D_n^{j}(a) := \frac {1}{2}\int_0^1dx\;j^2(x)+\sum_{k=1}^n
\int_0^1dx\;j(x)j(x_k)\;,\;n>0
\quad , \label{eq:gka1}
\ee
with $D_0(a)$ given by Eq.\ (\ref{eq:dkrw1}). Obviously, this series cannot
converge to the exact $D(a)$.

(2) By using the exact invariant density in the averages of Eq.(\ref{eq:gk1})
we define
\be
D_n^{\rho}(a) := \frac {1}{2}\int_0^1dx\;\rho_a^*(x) j^2_a(x)+\sum_{k=1}^n
\int_0^1dx\;\rho_a^*(x)j_a(x)j_a(x_k)\;,\;n>0
\quad , \label{eq:gka2}
\ee
here with $D_0^{\rho}(a)=\frac {1}{2}\int_0^1dx\;\rho_a^*(x) j^2_a(x)$, which
of course must converge exactly.  The approximations $D_n^{j}(a)$ and
$D_n^{\rho}(a)$ may be understood as time-dependent diffusion coefficients
according to the Green-Kubo formula Eq.\ (\ref{eq:gk1}). According to their
definitions, $D_n^{j}(a)$ enables us to look at contributions coming from
$j(x_n)$ only, whereas $D_n^{\rho}(a)$ assesses the importance of
contributions resulting from $\rho_a^*(x)$. The rates of convergence of both
approximations give an estimate of how important higher-order correlations are
in the different parameter regions of the diffusion coefficient $D(a)$.

Similarly to Eq.\ ({\ref{eq:dkrw2}), the next order $D_1(a)$ can easily be
calculated analytically and reads
\be
D_1(a)  = \left\{
\begin{array}{r@{\quad,\quad}l}
(a-2)/(2a) & 2< a\le1+\sqrt{3} \quad \mbox{and}\quad 2+\sqrt{2}<a\le4\\
3/2-3/a-2/a^2 & 1+\sqrt{3}<a\le3 \\
-1/2+3/a-2/a^2 & 3<a\le2+\sqrt{2}
\end{array}
\right\}  \label{eq:d1} \quad .
\ee
Further corrections up to order $n=3$ were obtained from computer simulations,
that is, an ensemble of point particles was iterated numerically according to
Eq.\ (\ref{eq:eom}). All results are contained in Fig.\ \ref{fig1} showing
that this hierarchy of correlated random walks generates a self-affine
structure which resembles, to some extent, the one of the well-known Koch
curve. Fig.\ \ref{fig1} (a) illustrates that this structure forms an important
ingredience of the exact diffusion coefficient $D(a)$ thus explaining basic
features of its fractality. Indeed, a suitable generalization of this approach
in the limit of time to infinity leads to the formulation of $D(a)$ in terms
of fractal generalized Takagi functions \cite{RKdiss,GaKl}.

Fig.\ \ref{fig1} (b) depicts the results for the series of $D_n^{\rho}(a)$ up
to order $n=2$ as obtained from computer simulations. This figure illustrates
that there exists a second source for an irregular structure related to the
integration over the invariant density, as was explained above. In the
Green-Kubo formula Eq.(\ref{eq:gk1}) both contributions are intimately coupled
with each other via the integration over phase space.

We now perform a more detailed analysis to reveal the precise origin of the
hierarchy of peaks in Fig.\ \ref{fig1}. For this purpose we redefine Eq.\
(\ref{eq:gka1}) as
\be
D_n^{j}(a) = \int_0^1dx\;j(x)J_n(x)-\frac
{1}{2}\int_0^1dx\;j^2(x)\;,\;n>0\quad , \label{eq:dkj}
\ee
where we have introduced the {\em jump velocity function}
\be
J_n(x):=\sum_{k=0}^n j(x_k) \quad . \label{eq:jvf}
\ee
From Eq.\ ({\ref{eq:jv}) it follows $J_n(x)=[x_{n+1}]$, that is, this function
gives the integer value of the displacement of a particle starting at some
initial position $x$. In Fig.\ \ref{fig2} we depict some representative
results for $J_1(x)$ under variation of the control parameter $a$. Because of
the symmetry of the map we restrict our considerations to $0<x<0.5$. Eq.\
(\ref{eq:dkj}) tells us that the product of this function with $j(x)$
determines the diffusion coefficient $D(a)$. The shaded bar in Fig.\
\ref{fig2} marks the subinterval  in which $j(x)=1$, whereas $j(x)=0$ otherwise,
thus an integration over $J_1(x)$ on this subinterval yields the respective
part of the diffusion coefficient. One can now relate the four diagrams (a) to
(d) in Fig.\ \ref{fig2} to the functional form of $D_1^{j}(a)$ in Fig.\
\ref{fig1} (a) thus understanding where the large peak in $D_1^{j}(a)$ for
$2.732<a<3.414$ comes from: For $a<2.732$, $J_1(x)$ does not change its
structure and the interval where particles escape to other unit intervals
increases monotonously, therefore $D(a)$ increases smoothly.  However,
starting from $a=2.732$ particles can jump for the first time to next nearest
neighbors within two time steps, as is visible in $J_1(x)$ taking values of
$2$ for $x$ close to $0.5$. Consequently, the slope of $D(a)$ increases
drastically leading to the first large peak around $a=3$. Precisely at $a=3$,
backscattering sets in meaning that particles starting around $x=0.5$ jump
back to the original unit interval within two time steps, as is reminiscent in
$J_1(x)$ in form of the region $J_1(x)=0$ for $x$ close to $0.5$. This leads
to the negative slope in $D(a)$ above $a=3$. Finally, particles starting
around $x=0.5$ jump to nearest neighbor unit intervals by staying there during
the second time step instead of jumping back. This yields again a monotonously
increasing $D(a)$ for $a>3.414$. Any higher-order peak for $D_n^{j}(a),n>1\;,$
follows from analogous arguments. Thus, the source of this type of fractality
in the diffusion coefficient is clearly identified in terms of the topological
instability of the function $J_1(x)$ under variation of the control parameter
$a$. Indeed, this argument not only quantifies two previous heuristic
interpretations of the structure of $D(a)$ as outlined in Refs.\
\cite{RKD,RKdiss,KlDo99}, it also explains why, on a very fine scale, there are
still deviations between these results and the precise location of the extrema
in $D(a)$, cp.\ to the ``overhang'' at $a=3$ as an example. The obvious reason
is that contributions from the invariant density slightly modifying this
structure are not taken into account.

Looking at the quantities $J_n(x)$ furthermore helps us to learn about the
rates of convergence of the approximations $D_n^{j}(a)$ to $D(a)$ at fixed
values of $a$, as is illustrated in Fig.\ \ref{fig3} (a) to (d). Here we have
numerically calculated $J_n(x)$ at $a=3.8$ for $n=0,1,2,3$. Again, the shaded
bar indicates the region where $j(x)=1$ enabling $J_n(x)$ to contribute to the
value of the diffusion coefficient according to Eq.\ (\ref{eq:dkj}). In fact,
$J_n(x)$ may also be interpreted as the {\em scattering function} of an
ensemble of particles starting from the unit interval, since it sensitively
measures the final position to which a particle moves within $n$ time steps
under variation of its initial position $x$. One can clearly see that, with
larger $n$, $J_n(x)$ develops more and more discontinuities eventually leading
to a highly singular and irregular function of $x$. Integration over further
and further refinements of $J_n(x)$ determines the convergence of the series
of $D_n^{j}(a)$ to a fixed value $D_{\infty}^{j}(a)$, cp.\ to Fig.\ \ref{fig1}
(a).

We remark that a suitable integration over the functions $J_n(x)$ leads to the
definition of fractal so-called generalized Takagi functions, which can be
calculated in terms of de Rham-type functional recursion relations
\cite{RKdiss,GaKl}. In a way, the integration over jump velocity functions
such as the one shown in Fig.\ \ref{fig3} (d) is similar to the integration
over Cantor set structures leading to Devils' staircase-type functions
\cite{Man82}. Our results presented so far thus bridge the gap between
understanding the coarse functional behavior of $D(a)$ on the basis of simple
random walk approximations only, and analyzing its full fractal structure in
terms of Takagi-like fractal forms, in combination with an integration over a
complicated non-uniform invariant density. We now show that essentially the
same line of argument can be successfully applied to more physical dynamical
systems such as particle billiards.

\section{Deterministic diffusion in billiards}\label{sect:plg}
The class of two-dimensional billiards we want to discuss here is described as
follows: A point particle undergoes elastic collisions with obstacles of the
same size and shape whose centers are fixed on a triangular lattice. There is
no external field, thus the equations of motion are defined by the Hamiltonian
$H=mv^2/2$ supplemented by geometric boundary conditions as induced by the
scatterers. A standard example is the periodic Lorentz gas for which the
scatterers consist of hard disks of radius $R$, see Fig.\ \ref{fig4}
\cite{Do99,Gasp}. In the following we choose $m=1$, $v=1$, $R=1$, and as a
control parameter we introduce the smallest inter disk distance $w$ such that
the lattice spacing of the disks is $2+w$. $w$ is related to the number
density $n$ of the disks by $n(w)=2/[\sqrt{3}(2+w)^2]$. At close packing $w=0$
the moving particle is trapped in a single triangular region formed between
three disks, see Fig.\ \ref{fig4}, part (1). For
$0<w<w_{\infty}=4/\sqrt{3}-2=0.3094$, the particle can move across the entire
lattice, but it cannot move collision-free for an infinite time. For
$w>w_{\infty}$ the particle can move ballistic-like in form of arbitrarily far
jumps between two collisions.

The diffusion coefficient for this particle billiard can be defined by the
two-dimensional equivalent of the Einstein formula Eq.(\ref{eq:dkdef1})
reading
\be
D(w)=\lim_{t\to\infty}<({\bf x}(t)-{\bf x}(0))^2>/(4t)\quad , \label{eq:dkdef2}
\ee
where, again, the average is taken over the equilibrium distribution of
particles with position coordinates ${\bf x}(t)$. It can be proven that in the
regime of $0<w<w_{\infty}$ the parameter-dependent diffusion coefficient
$D(w)$ exists \cite{BuSi80a}. The full parameter-dependence of this function
was discussed particularly in Ref.\ \cite{KlDe00} showing that, on a fine
scale, $D(w)$ is again an irregular function of the parameter $w$ similarly to
the diffusion coefficient of the one-dmensional map $D(a)$ as discussed
above. Whether $D(w)$ is indeed fractal, or maybe $C^1$ but not $C^2$ in
contrast to the one-dimensional map discussed above, is currently an open
question. The main issue we want to focus on in this section are quantitative
approximations for the full parameter dependence of $D(w)$, and to check for
the importance of memory effects. A first simple analytical approximation for
the diffusion coefficient was derived by Machta and Zwanzig in Ref.\
\cite{MaZw83}. This solution was based on the assumption that diffusion can be
treated as a Markovian hopping process between the triangular trapping regions
indicated in Fig.\ \ref{fig4}. For random walks on two-dimensional isotropic
lattices the diffusion coefficient then reads $D=\ell^2/(4\tau)$, where
$\ell=(2+w)/\sqrt{3}$ is the smallest distance between two centers of the
traps, and $\tau^{-1}$ is the average rate at which a particle leaves a
trap. This rate can be calculated by the fraction of phase space volume being
available for leaving the trap divided by the total phase space volume of the
trap thus leading to the Machta-Zwanzig random walk approximation of the
diffusion coefficient
\be
D_{\rm MZ}(w)=\frac{w(2+w)^2}{\pi[\sqrt{3}(2+w)^2-2\pi]} \quad . \label{eq:dmz}
\ee
Indeed, this approximation is precisely the billiard analogue to the
one-dimensional random walk diffusion coefficient for maps
Eqs.(\ref{eq:dkrw1}),(\ref{eq:dkrw2}). Similarly, this approximation is
asymptotically exact only for $w\to0$, as is shown in comparison to computer
simulation results in Fig.\ \ref{fig5} \cite{KlDe00,MaZw83}.

In Ref.\ \cite{KlDe00}, Eq.\ (\ref{eq:dmz}) was systematically improved by
including higher-order correlations. Two basic approaches were presented both
starting from the idea of Machta and Zwanzig of looking at diffusion in the
Lorentz gas as a hopping process on a hexagonal lattice of ``traps'' with
frequency $\tau^{-1}$. This picture was quantified by introducing a simple
symbolic dynamics for a particle moving from trap to trap as indicated in
Fig.\ \ref{fig4}. Let us follow a long trajectory of a particle starting with
velocity ${\bf v}$ parallel to the $y$-axis, cp.\ part (1) in Fig.\
\ref{fig4}. For each visited trap we label the entrance through which the
particle entered with $z$, the exit to the left of this entrance with $l$, and
the one to the right with $r$, see part (2) in Fig.\ \ref{fig4}. Thus, a
trajectory in the Lorentz gas is mapped onto words composed of the alphabet
$\{z, l,r\}$.  One can now associate transition probabilities to these symbol
sequences reading, for a time interval of $2\tau$, $p(z), p(l), p(r)$. $p(z)$
corresponds to the probability of backscattering, whereas
$p(l)=p(r)=(1-p(z))/2$ indicates forward scattering. For a time interval of
$3\tau$, we have nine symbol sequences each consisting of two symbols leading
to the probabilities $p(zz)$, $p(zl)$, $p(zr)$, $p(lz)$, $p(ll)$, $p(lr)$,
$p(rz)$, $p(rl)$, $p(rr)$, and so on. Beside this hierarchy of conditional
probabilities defined on a symbolic dynamics there is a different type of
probability that a particle leaving a trap jumps without any collision
directly to the next nearest neighbor trap. As was shown in Ref.\
\cite{KlDe00}, Eq.\ (\ref{eq:dmz}) can be corrected by analytically including
all these probabilities. Alternatively, lattice gas computer simulations were
performed by using the probabilities as associated to the symbol
sequences. The heuristic corrections of Eq.\ (\ref{eq:dmz}) led to a
satisfactorily explanation of the overall behavior of $D(w)$ on a coarse
scale, however, the convergence was not exact. The lattice gas simulations, on
the other hand, were converging exactly, however, here a proper analytical
expression for the diffusion coefficient approximations in terms of the
associated probabilities was not available.

In analogy to the procedure as outlined for the one-dimensional map, that is,
starting from a suitable Green-Kubo formula, we will now define a third
approximation scheme which we expect to be generally applicable to diffusion
in particle billiards. Compared to the two existing approaches mentioned above
the advantage of the new method is two-fold, namely (1) that by using the set
of symbolic probabilities the respective Green-Kubo formula can be evaluated
according to an analytical scheme, and (2) that the resulting approximations
converge exactly to the computer simulation results.

In the Appendix we prove that, starting from the Einstein formula Eq.\
(\ref{eq:dkdef2}), quite in analogy to the one-dimensional case a Green-Kubo
formula can be derived which is defined for an ensemble of particles moving on
the hexagonal lattice of traps depicted in Fig.\ \ref{fig4}. The result reads
\be
D(w) = \frac {1}{4\tau}\left\langle {\bf j}^2({\bf x}_0)\right\rangle
+\frac{1}{2\tau}\sum_{n=1}^\infty\left\langle {\bf j}({\bf x}_0)\cdot{\bf
j}({\bf x}_n)\right\rangle
\quad . \label{eq:gk2}
\ee
Here ${\bf j}({\bf x}_n)$ defines jumps at the $n$th time step in terms of the
lattice vectors $\bml{\ell}_{\alpha\beta\gamma\ldots}$ associated to the
respective symbol sequence of the full trajectory on the hexagonal lattice,
cp.\ to Fig.\ \ref{fig4}. Let us start with ${\bf j}({\bf
x}_0)=\bml{\ell}/\tau\;,\;\bml{\ell}:=(0,\ell)^*$. The next jumps are then
defined by ${\bf j}({\bf
x}_1)=\bml{\ell}_{\alpha}/\tau\;,\;\alpha\in\{l,r,z\}$, and so on. The
averages indicated in Eq.\ (\ref{eq:gk2}) by the brackets are calculated by
weighting the respective scalar products of lattice vectors with the
corresponding conditional probablities $p(\alpha\beta\gamma\ldots)$. Note that
Eq.\ (\ref{eq:gk2}) is the honeycomb lattice analogue to the Green-Kubo
formula derived by Gaspard for the Poincar\'e-Birkhoff map of the periodic
Lorentz gas \cite{Gasp,Gasp96}. The Poincar\'e-Birkhoff version is very
efficient for numerical computations, however, according to its construction
it fails to reproduce the Machta-Zwanzig approximation Eq.\
(\ref{eq:dmz}). Consequently, it does not appear to be very suitable for
diffusion coefficient approximations of low order. More details will be
discussed elsewhere \cite{TH01}. We remark that, in terms of using a symbolic
dynamics, there is also some link between Eq.\ (\ref{eq:gk2}) and respective
diffusion coefficient formulas obtained from periodic orbit theory
\cite{CEG91,CvGS92}.

We now demonstrate how Eq.\ (\ref{eq:gk2}) can be used for systematic
improvements of the diffusion coefficient on the lattice of traps by including
dynamical correlations: As in case of one-dimensional maps we start by looking
at the first term in Eq.\ (\ref{eq:gk2}) and cut off all higher order
contributions. Obviously, the first term is again the random walk expression
for the diffusion coefficient on the hexagonal lattice of traps which, by
including the respective solution for the jump frequency $\tau^{-1}$, boils
down to Eq.\ (\ref{eq:dmz}). For calculating higher-order corrections we now
define the hierarchy of approximations
\be
D_n(w) =
\frac{l^2}{4\tau}+\frac{1}{2\tau}\sum_{\alpha\beta\gamma\ldots}p(\alpha\beta\gamma\ldots)
\bml{\ell}\cdot\bml{\ell}(\alpha\beta\gamma\ldots)\;,\;n>0 \quad , \label{eq:gka3}
\ee
with $D_0(w)$ given by Eq.\ (\ref{eq:dmz}). To our knowledge yet there is no
method available to analytically calculate the conditional probabilities
$p(\alpha\beta\gamma\ldots)$. Our following evaluations are therefore based on
the data presented in Ref.\ \cite{KlDe00} as obtained from computer
simulations. In terms of the formal probabilities it is now easy to calculate
the solution for the first order approximation at time step $2\tau$ to
\be
D_1(w)=D_0(w)+D_0(w)(1-3p(z)) \label{eq:dbs} \quad .
\ee
For a comparison of this formula with the simulation data $D(w)$ see Fig.\
\ref{fig5}. We remark that the corresponding solution in Ref.\ \cite{KlDe00}
as obtained from a heuristical correction of Eq.\ (\ref{eq:dmz}) reads
$D_{1,{\rm MZ}}(w)=D_0(w)3(1-p(z))/2$. Indeed, one can show that the new
formula Eq.(\ref{eq:dbs}) is closer to $D(w)$ for large $w$, whereas for small
$w$ the previous approximation is somewhat better.  It is straightforward to
calculate the two approximations of next highest order to
\be
D_2(w)=D_1(w)+D_0(w)\left(2p(zz)+4p(lr)-2p(ll)-4p(lz)\right)
\ee
and to
\bna
D_3(w)&=&D_2(w)+D_0(w)\left[p(llr)+p(llz)+p(lrl)+p(lrr)+p(lzl)+p(lzz)+p(rll)+p(rlr)+p(rrl)\right.
\\ \nonumber
 & &+p(rrz)+p(rzr)+p(rzz)+p(zll)+p(zlz)+p(zrr)+p(zrz)+p(zzl)+p(zzr)\\ \nonumber
 & &\left. -2(p(lll)+p(lrz)+p(lzr)+p(rlz)+p(rrr)+p(rzl)+p(zlr)+p(zrl)+p(zzz))\right]
\quad .
\ena
All results are shown in Fig.\ \ref{fig5} demonstrating that the series of
approximations defined by Eq.\ (\ref{eq:gka3}) converges quickly and
everywhere to the simulation results. Our new scheme thus eliminates the
deficiency of the semi-analytical approximation proposed in Ref.\
\cite{KlDe00} that was based on heuristically correcting the
Machta-Zwanzig approximation Eq.\ (\ref{eq:dmz}). By comparing this new scheme
with the lattice gas simulations of the same reference, on the other hand, it
turns out that the rate of convergence of the lattice gas approach is still a
bit better. In any case, all these three methods unambiguously demonstrate
that for achieving a complete understanding of the density-dependent diffusion
coefficient in the periodic Lorentz gas it is unavoidable to take higher-order
correlations, or the impact of memory effects, properly into account.

We finally remark that a very good low-order approximation for the diffusion
coefficient can already be obtained by combining Eq.\ (\ref{eq:dbs}) with the
probability of collisionless flights $p_{\rm cf}(w)$ mentioned above, that is,
by taking into account the possibility of next nearest neighbor jumps. The
correction of $D_0(w)$ as given by Eq.\ (\ref{eq:dmz}) according to
collisionless flights only was already calculated in Ref.\ \cite{KlDe00} and
read $D_{0,{\rm cf}}(w)=D_0(w)(1+2p_{\rm cf}(w))$. Adding now the second term
of Eq.\ (\ref{eq:dbs}) to this expression by just following the Green-Kubo
scheme yields
\be
D_{1,{\rm cf}}(w)=D_0(w)(2+2p_{\rm cf}(w)-3p(z)) \quad . \label{eq:dcf}
\ee
This solution is also depicted in Fig.\ \ref{fig5} and shows that this
approximation indeed significantly improves Eq.\ (\ref{eq:dbs}) for large $w$
yielding a function that is qualitatively and quantitatively very close to
$D(w)$. We know of no better approximation for $D(w)$ based on information
such as $p(z)$ and $p_{\rm cf}(w)$ only. The succesful application of Eq.\
(\ref{eq:dcf}) suggests that collisionless flights form an important mechanism
to understand the full diffusive dynamics of this billiard. However, somewhat
surprisingly they are not explicitly contained neither in the Green-Kubo
expansion Eq.\ (\ref{eq:gka3}) nor in the lattice gas simulations of Ref.\
\cite{KlDe00}. In both cases exact convergence is achieved by following the
hierarchy of symbol sequence probabilities only, where collisionless flights
are not apparent.

\section{Summary and conclusions}
In this paper we suggested a general scheme of how to understand the structure
of parameter-dependent deterministic diffusion coefficients in simple model
systems. The important point was to find suitable Green-Kubo formulas to start
with. We used the fact that the class of models we studied here was defined on
periodic lattices, and respectively we discretized the dynamics of the moving
particles according to these lattices. As two different examples we analyzed a
simple one-dimensional map on the line as well as the periodic Lorentz gas. In
both cases we recovered the respective well-known random walk formulas for the
diffusion coefficient as zero-order approximations in our Green-Kubo
approach. We then calculated higher-order terms according to Green-Kubo thus
systematically including higher-order correlations. As much as possible this
was performed analytically, alternatively in combination with data obtained
from computer simulations. Our results provided clear evidence that a proper
understanding of the parameter-dependent diffusion coefficients in both models
can only be achieved by taking strong memory effects into account. In this
respect our research appears to be somewhat related to the findings of
long-time tails in the velocity autocorrelation function of simple model
systems such as Lorentz gases, and to the existence of non-analyticities in
the density expansion of transport coefficients, which are both consequences
of strong correlations in the dynamics of the moving particles; see Ref.\
\cite{DoBe77} for a nice review. In case of the one-dimensional map our
approach enables a detailed understanding of the dynamical mechanism
generating the most pronounced irregularities in the fractal diffusion
coefficient. In case of the periodic Lorentz gas this scheme straightforwardly
generalizes the Machta-Zwanzig random walk formula in terms of systematic
higher-order approximations that converge exactly to the simulation results.

For one-dimensional maps the infinite time limit of this approach, though not
the intermediate level as quantitatively discussed here, was already worked
out in Refs.\ \cite{RKdiss,GaKl} leading to the definition of the diffusion
coefficient in terms of fractal so-called generalized Takagi functions. For
the periodic Lorentz gas an analogous generalization would be desirable as
well.  We furthermore remark that the approximation scheme as presented in
this paper was already successfully applied (1) to the one-dimensional
climbing sine map, where there is a complicated transition scenario between
normal and anomalous diffusion \cite{KoKl01}, and (2) to the so-called flower
shape billiard, where the hard disks of the Lorentz gas are replaced by
obstacles of flower shape \cite{HKG01}. In both cases the resulting
parameter-dependent diffusion coefficients, as far as existent, are much more
complicated functions than in the corresponding models discussed above, yet
our scheme yields systematic explanations of the structure of these functions
in terms of strong dynamical correlations. We thus expect that using
Green-Kubo formulas this way provides a general access road to understanding
deterministic diffusion in low-dimensional periodic arrays of scatterers,
possibly also in view of experimental results on systems such as antidot
lattices, ratchets, and Josephson junctions.\\

\noindent {\bf Acknowledgments}\\
We thank T.Harayama for interesting discussions, L.Matyas for a careful
reading of the manuscript, and Chr.Dellago for originally having generated the
data set that was used for the Lorentz gas.

\begin{appendix}

\section{Green-Kubo formula on the hexagonal lattice}\label{app}
In this Section we derive Eq.\ (\ref{eq:gk2}) which yields the diffusion
coefficient $D(w)$ for the periodic Lorentz gas via a generalization of the
Machta-Zwanzig picture. That is, we look at diffusion as a higher-order Markov
process on a hexagonal lattice of traps with lattice spacing $\bml{\ell}$
where particles hop with frequency $\tau^{-1}$ from trap to trap, cp.\ to
Fig.\ \ref{fig4}. The time $t$ is suitably rewritten in terms of the escape
time $\tau=(\pi/6w)(\sqrt{3}/2(2+w)^2-\pi)$
\cite{MaZw83} as $t=n\tau\;,\;n\in {\mb N}$. Let ${\bf x}_n$
be the position of the moving particle at time step $n$. We then write ${\bf
x}_n\equiv{\bf X}_n+{\bf \tl{x}}_n$, where ${\bf X}_n$ denotes the position of
the trap on the hexagonal lattice in which the particle is situated at time
step $n$. This vector can be expressed by a suitable combination of lattice
vectors. For example, one may choose a sum of
$\bml{\ell}_{\alpha\beta\gamma\ldots}$ as introduced in Section
\ref{sect:plg}, see Fig.\ \ref{fig4}, a more precise definition is
not important here. Correspondingly, ${\bf \tl{x}}_n$ is the distance from the
nearest trap center to the actual position of the particle in the Wigner-Seitz
cell.

The Einstein formula Eq.\ (\ref{eq:dkdef2}) then reads
\be
D(w)=\lim_{n\to\infty}\left<({\bf X}_n+{\bf \Delta\tl{x}}_n)^2\right>/(4n\tau) \label{eq:dkdefa}
\ee
where ${\bf \Delta \tl{x}}_n:=\bml{\tl{x}}_n-\bml{\tl{x}}_0$. Multiplying out
the nominator we get
\be
<{\bf X}_n^2+2{\bf X}_n\cdot{\bf \Delta \tl{x}}_n+{\bf \Delta
\tl{x}}_n^2>\quad \label{eq:mult}
\ee
According to its definition the last term is bounded, ${\bf \Delta
\tl{x}}_n^2<const$. To the second term we apply the H\"older inequality \cite{LaMa}
yielding
\be
|<{\bf X}_n\cdot{\bf \Delta \tl{x}}_n>|\le \sqrt{<|{\bf
X}_n|^2>}\sqrt{<|{\bf \Delta
\tl{x}}_n|^2>}<const \sqrt{<|{\bf X}_n|^2>} \quad .
\ee
Consequently, in the limit of infinite time only the first term in Eq.\
(\ref{eq:mult}) contributes to the positive diffusion coefficient $D(w)$ of
Eq.\ (\ref{eq:dkdefa}) leading to
\be
D(w)=\lim_{n\to\infty}<{\bf X}_n^2>/(4n\tau) \quad . \label{eq:dkx}
\ee
Starting from this Einstein formula on the hexagonal lattice it is now
straightforward to derive Eq.\ (\ref{eq:gk2}). ${\bf X}_n^2=X_n^2+Y_n^2$ tells
us that essentially there are two one-dimensional parts, thus we are back to
the respective derivation of Eq.\ (\ref{eq:gk1}) in Ref.\
\cite{Do99}. To make this precise, let us write
\be
X_n=\sum_{k=0}^{n-1}j(x_k)\quad , \label{eq:jsum}
\ee
which is in analogy to Eq.\ (\ref{eq:jvf}). However, here
$j(x_k):=X_{k+1}-X_k$ is strictly speaking no jump {\em velocity}, since it
only denotes the distance a particle jumps within a time interval $\tau$,
whereas $\tau=1$ in Eq.\ (\ref{eq:jv}). Multiplying out the nominator of Eq.\
(\ref{eq:dkx}) in terms of Eq.\ (\ref{eq:jsum}) leads to an equation for
$D(w)$ in form of velocity autocorrelation functions $C(k,l):=<j(x_k)\cdot
j(x_l)>=\int dx\;dy\;\rho^*(x,y)\;j(x_k)\cdot j(x_l)$, where $\rho^*(x,y)$ is
the equilibrium density of the periodic Lorentz gas. Translational invariance
implies $C(k,l)=C(0,l-k)$ which is easily shown by substitution combined with
conservation of probability according to the Frobenius-Perron equation of the
billiard. Summing up all contributions $C(0,k)$ obtained from the
multiplication, doing the same for the component $Y_n$, and putting both
results together yields Eq.\ (\ref{eq:gk2}).

\end{appendix}

\newpage
\begin{figure}
\begin{center}
\epsfxsize=11cm
\centerline{\rotate[r]{\epsfbox{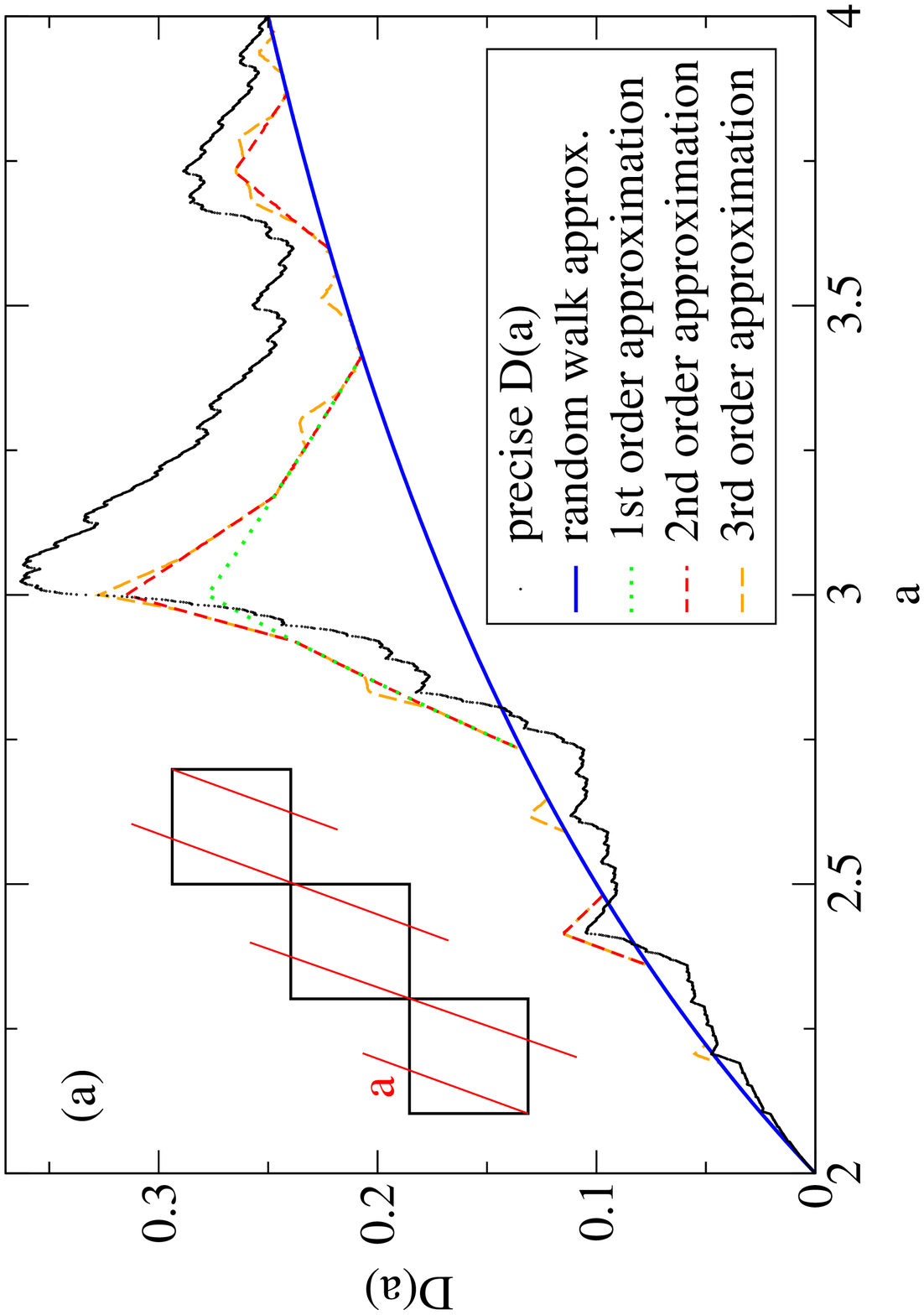}}}
  
\vspace*{-1.5cm}
\epsfxsize=11cm
\centerline{\rotate[r]{\epsfbox{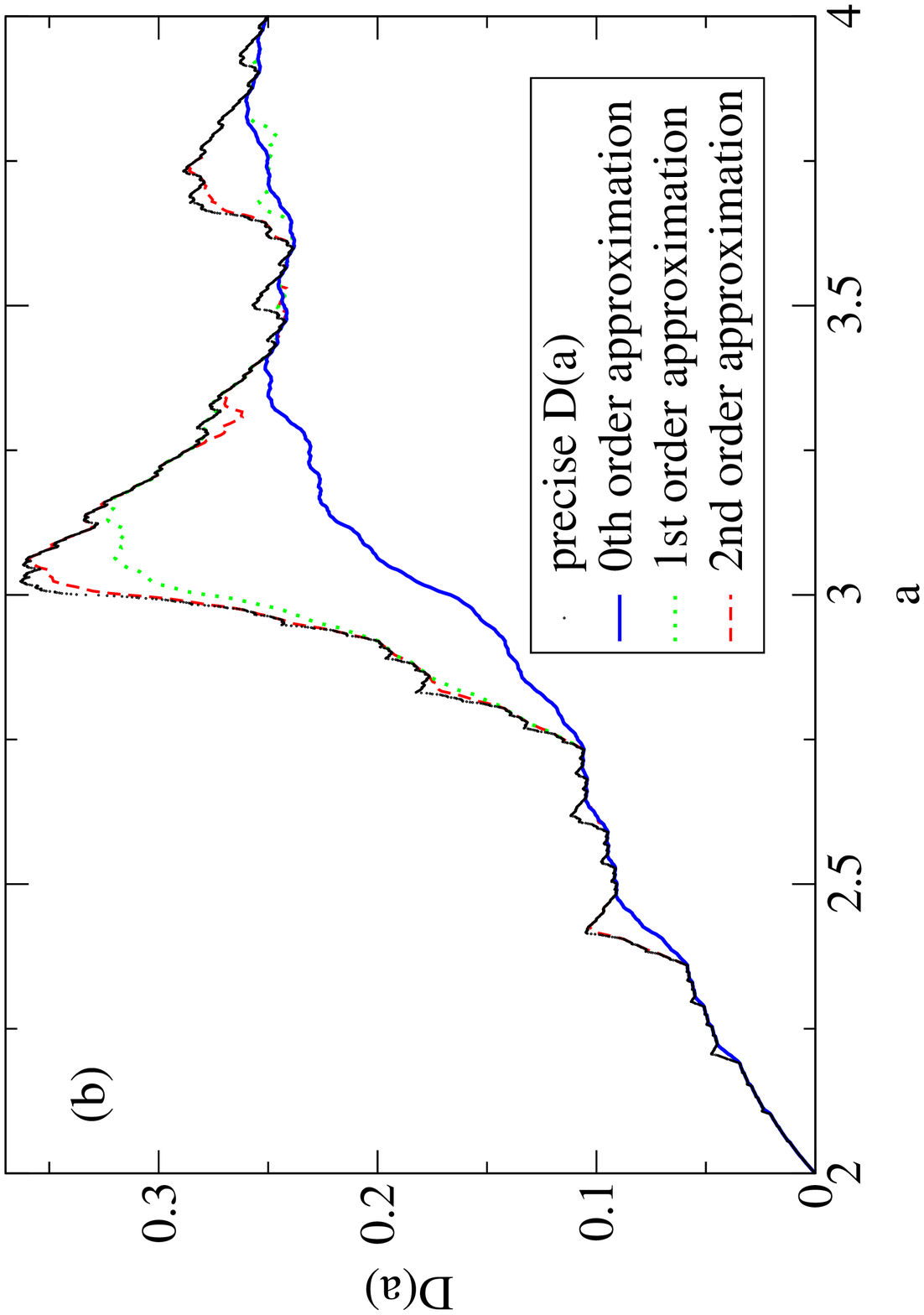}}}
\end{center}
\vspace*{-0.5cm}
\caption{Diffusion coefficient $D(a)$ for the one-dimensional map shown in the upper
left part of (a), where $a$ is the slope of the map. The dots are obtained
from the method of Refs.\ [2--4], the different lines correspond to different
levels of approximations based on the Green-Kubo formula Eq.\
(\ref{eq:gk1}). In (a) the approximations were computed from Eq.\
(\ref{eq:gka1}) assuming a constant invariant density, in (b) they are from
Eq.\ (\ref{eq:gka2}) which includes the exact invariant density. Any error
bars are smaller than visible. All quantities here and in the following
figures are without units.}
\label{fig1}
\end{figure}

\begin{figure}
\begin{center}
\epsfxsize=15cm
\centerline{\rotate[r]{\epsfbox{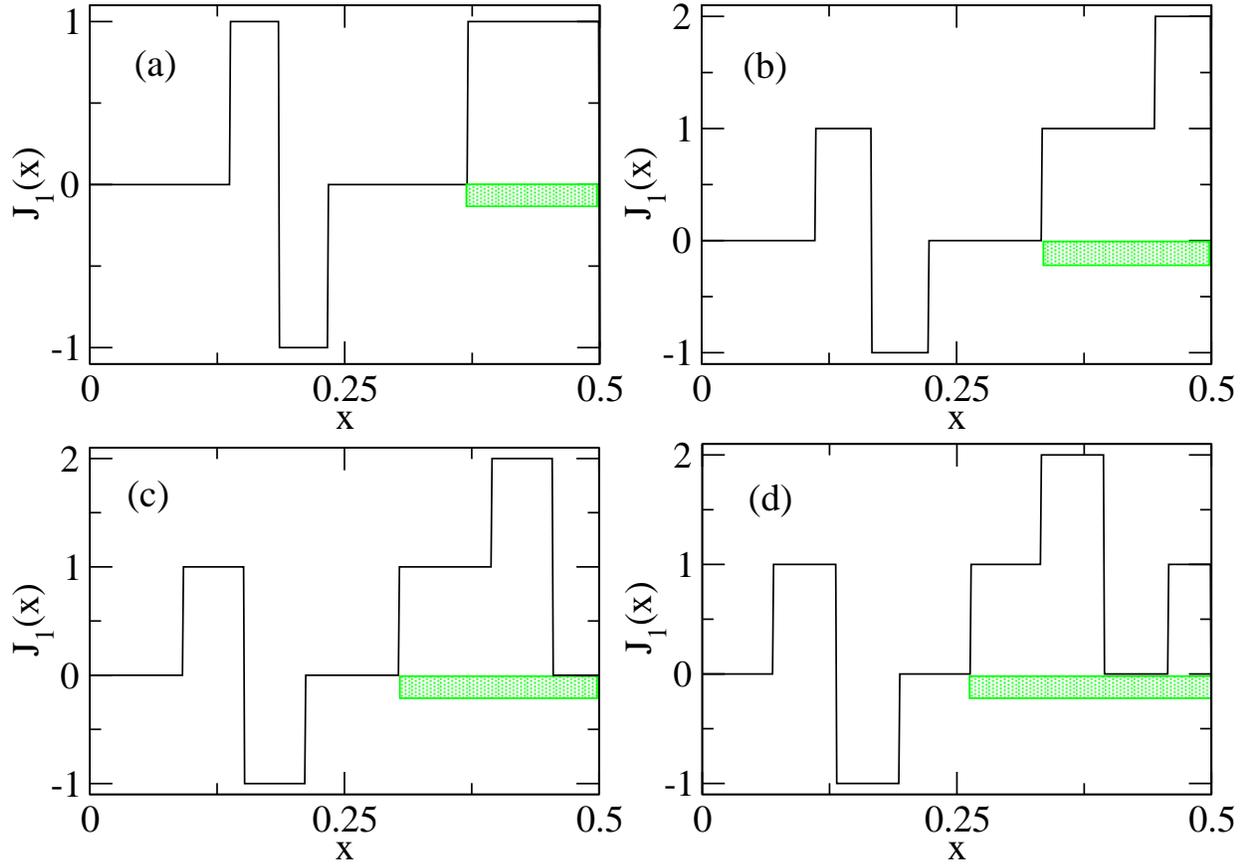}}}
\end{center}
\caption{Jump velocity function $J_1(x)$ as defined by Eq.\ (\ref{eq:jvf})
which gives the integer value of the displacement of a particle starting at
some initial position $x$. Shown are results for different values of the slope
$a$. It is (a) $a=2.7$, (b) $a=3.0$, (c) $a=3.3$, (d) $a=3.8$. The shaded bar
marks the subinterval where the jump velocity $j(x_n)$ defined in Eq.\
(\ref{eq:jv}) is equal to one.}
\label{fig2}
\end{figure}

\begin{figure}
\begin{center}
\epsfxsize=15cm
\centerline{\rotate[r]{\epsfbox{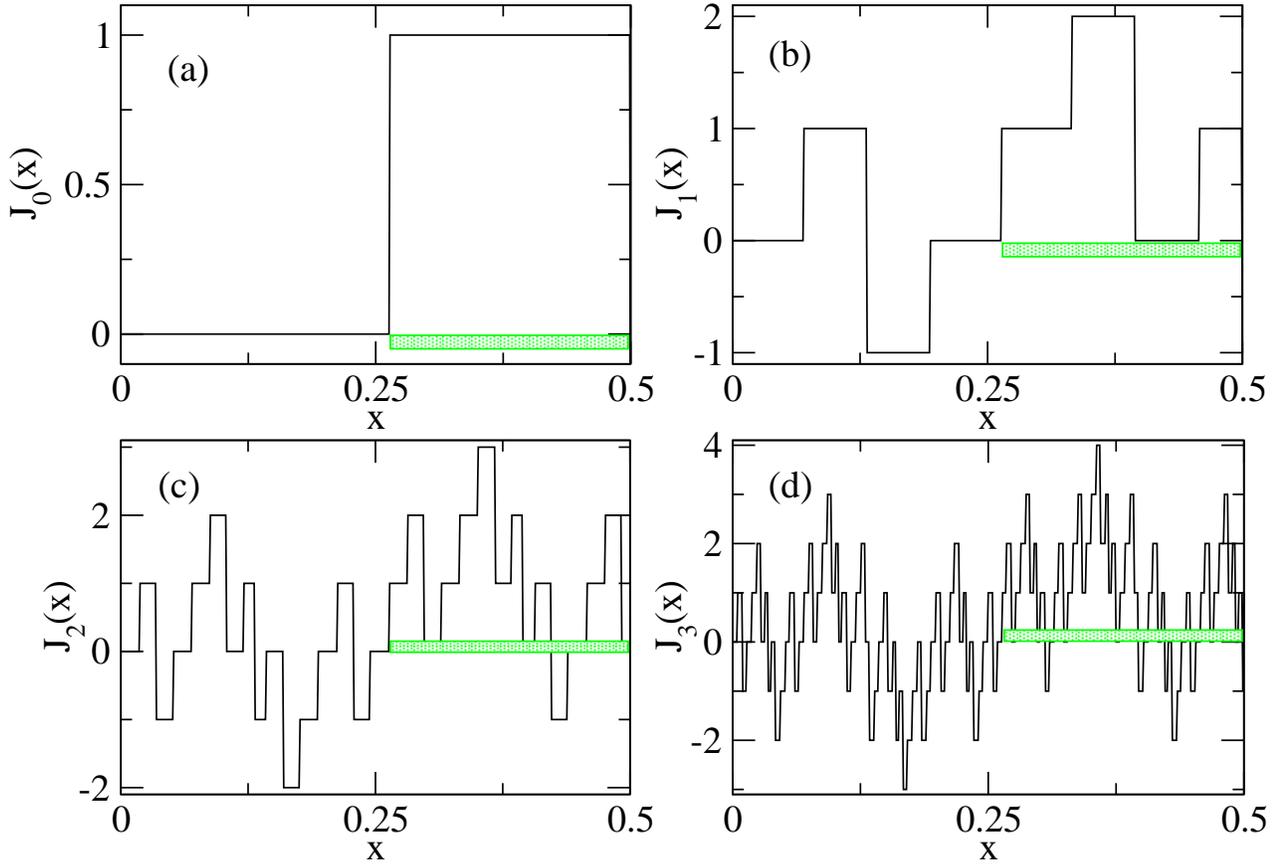}}}
\end{center}
\caption{Jump velocity function $J_n(x)$ as defined by Eq.\ (\ref{eq:jvf}) at
the fixed parameter value $a=3.8$ for the different number of time steps (a)
$n=0$, (b) $n=1$, (c) $n=2$, (d) $n=3$. Again, the shaded bar marks the
subinterval where the jump velocity $j(x_n)$ defined in Eq.\ (\ref{eq:jv}) is
equal to one.}
\label{fig3}
\end{figure}

\begin{figure}
\begin{center}
\epsfxsize=8cm
\centerline{\rotate[r]{\epsfbox{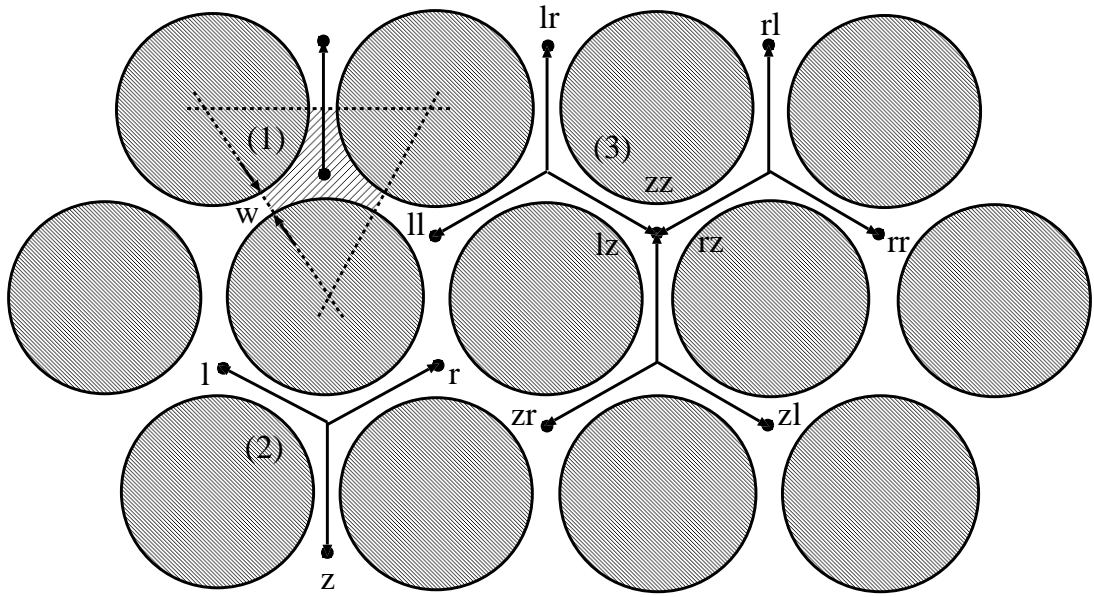}}}
\end{center}
\caption{Geometry of the periodic Lorentz gas with the gap size $w$
as control parameter. The hatched area related to (1) marks a so-called
trapping region, the arrow gives the lattice vector connecting the center of
this trap to the next one above. In (2) three lattice vectors are introduced
and labeled with the symbols $l,r,z$. They indicate the positions where
particles move along the hexagonal lattice of Wigner-Seitz cells, starting
from the trap $z$, within two time steps of length $\tau$. (3) depicts the
situation after three time steps $\tau$ with the different lattice vectors
associated to symbol sequences of length two.}
\label{fig4}
\end{figure}

\begin{figure}
\begin{center}
\epsfxsize=13cm
\centerline{\rotate[r]{\epsfbox{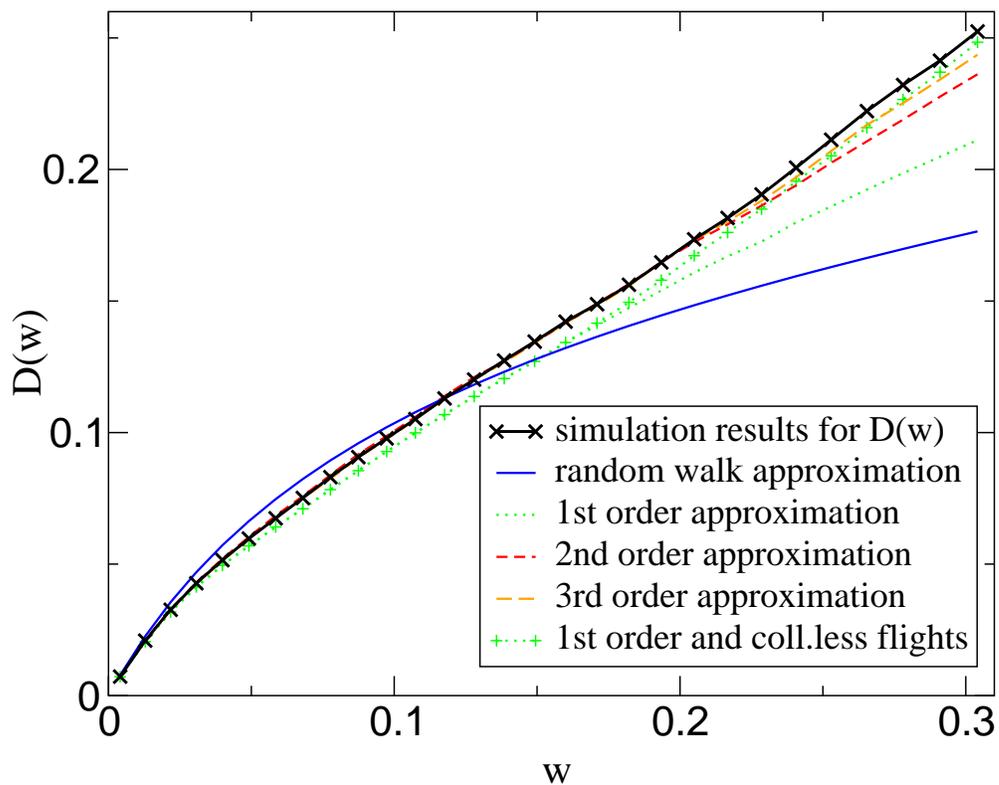}}}
\end{center}
\caption{Diffusion coefficient $D(w)$ for the periodic Lorentz gas as a
function of the gap size $w$ as a control parameter. The computer simulation
results for $D(w)$ are from Ref.\ [8], error bars are much smaller than the
size of the symbols. The other lines correspond to different levels of the
approximation Eq.\ (\ref{eq:gka3}), the last approximation (again with
symbols) is from Eq.\ (\ref{eq:dcf}).}
\label{fig5}
\end{figure}

\end{document}